\def\bq{\begin{equation}}
\def\eq{\end{equation}}
\def\bqa{\begin{eqnarray}}
\def\eqa{\end{eqnarray}}
\def\bqb{\begin{eqnarray*}}
\def\eqb{\end{eqnarray*}}
\documentstyle[12pt,epsf]{article}
\def\pr#1#2#3{ Phys. Rev. ${\bf{#1}}$ (#2) #3}
\def\prl#1#2#3{ Phys. Rev. Lett. ${\bf{#1}}$ (#2) #3}
\def\pl#1#2#3{ Phys. Lett. ${\bf{#1}}$ (#2) #3}

\def\np#1#2#3{ Nucl. Phys. ${\bf{#1}}$ (#2) #3}
\def\zp#1#2#3{ Z. Phys. ${\bf{#1}}$ (#2) #3}

\def\ie{{\it i.e.\/}}
\def\eg{{\it e.g.\/}}

\def\etal{{\it et.al.\/}}

\global\nulldelimiterspace = 0pt
 
\def\Bsl{\hbox{/\kern-.6700em$B$}} %  Bslash
\def\Dsl{\hbox{/\kern-.6700em$D$}} %  Dslash
\def\Wsl{\hbox{/\kern-.6700em$W$}} %  Wslash 

\def\roughly#1{\mathrel{\raise.3ex
    \hbox{$#1$\kern-.75em\lower1ex\hbox{$\sim$}}}}
\def\lsim{\roughly<}
\def\gsim{\roughly>}

\def\ol#1{\overline{#1}}

\def\L{ {\cal L }}
\def\O{ {\cal O }}

\def\mwd{M_W^2}
\def\mw{M_W}
\def\mt{m_t}
\def\mtd{m_t^2}

\def\lam{\Lambda_{NP}}

\begin{document}
\pagenumbering{arabic}
\thispagestyle{empty}
\def\thefootnote{\fnsymbol{footnote}}
\setcounter{footnote}{1}
 
\begin{flushright}
PM/96-28 \\ THES-TP 96/09 \\
revised version
%hep-ph/9609437 \\
%September 1996 
 \end{flushright}
\vspace{2cm}
%---------------------titre ---------------------------------------
\begin{center}
{\Large\bf General description of New Physics in $t$, $b$ and 
boson Interactions and its Unitarity 
Constraints}\footnote{Partially supported by the EC contract 
CHRX-CT94-0579.}
 \vspace{1.5cm}  \\
%-----------------------------------------------------------------
{\large G.J. Gounaris$^a$, D.T. Papadamou$^a$ and F.M. Renard$^b$}
\vspace {0.5cm}  \\

$^a$Department of Theoretical Physics, University of Thessaloniki,\\
Gr-54006, Thessaloniki, Greece.\\
\vspace{0.2cm}
$^b$Physique
Math\'{e}matique et Th\'{e}orique,
UPRES-A 5032\\
Universit\'{e} Montpellier II,
 F-34095 Montpellier Cedex 5.\\

\vspace {1cm}
 
{\bf Abstract}
\end{center}
We list all possible $dim=6$ CP conserving and $SU(3)\times
SU(2) \times U(1)$ gauge invariant interactions, which could 
be generated in case no new particles would be reachable in the
future Colliders, and the only observable New Physics 
would be in the form of new interactions affecting the scalar
sector and the quarks of the third family. These interactions
are described by operators involving the standard model scalar
field, the quarks of the third family and the gauge bosons.
Subsequently, we identify those operators which do not
contribute to LEP1 (and lower energy) observables at tree level
and are not purely gluonic.
Since present measurements do not strongly constrain 
the couplings of these operators, we derive here the 
~unitarity bounds on them.
Finally, in order to get a feeling on the possible physical 
meaning of the appearance of any of these operators, 
we identify the operators generated in a class of ~renormalizable
dynamical models which at the TeV scale, are fully described by 
the $SU(3)\times SU(2) \times U(1)$  gauge group.

\vspace{0.5cm}
\def\thefootnote{\arabic{footnote}}
\setcounter{footnote}{0}
\clearpage

\section{Introduction}

Up to now the Standard Model (SM) has passed all tests and stands in
an amazing agreement with the experimental data \cite{SM1, SM2}.
Minor discrepancies which are occasionally announced tend to
disappear as the statistics is increasing \cite{SM2}. 
Nevertheless it is
widely believed that there is some New Physics (NP) to be
learned which is beyond SM and which will help clarifying the
mysterious mechanism of the spontaneous gauge symmetry breaking.
In other words, it is quite commonly expected that the 
NP which may be discovered some day,  will be related 
to the way the Higgs particle(s) is generated and 
interacts \cite{NPgen, NP-GLR}.\par 

It may turn out that no scalar particles really exist and 
that the Higgs induced New Physics takes the form of a new strong
interaction among the longitudinal $W$ and 
$Z$ bosons \cite{Chano}. This possibility is widely studied
these days, but it
will not concern us here \cite{Appel, NPnonlin}. 
Instead, the present work is within the 
alternative option that one or
more Higgs particles exist having masses of the order
of the electroweak scale $v=(\sqrt{2} G_\mu)^{-1/2})=0.246~TeV$. 
One example of such a philosophy is of course SUSY, where  
the desire to invent a mechanism which "naturally" accommodates 
$m_H \sim v$, leads to the intriguing 
consequence that every known particle should have a
supersymmetric partner,
some of which must be reachable in the future and may be even
the present Colliders.\par

We should remark though, that even if NP remains perturbative, 
it may turn out that the 
underlying cause of spontaneous breaking is much more contrived 
than the present realizations of SUSY suggest. One viable
possibility for the future could be that 
no new particles will be reachable at the
future Colliders, apart of course from the theoretically well
known standard model
Higgs particle. If this turns out to be the case, then 
NP could only appear in the form of slight modifications of 
the SM interactions among
the known particles and of course the SM Higgs 
\cite{NPgen}. \par

How we could parameterize these NP interactions in a rather
general way, provided that no particles beyond
those already present in SM, will be
reachable in the future Colliders?
To achieve such a ~description, we subscribe to the idea that NP 
stems from the scalar sector and ~observe that 
the Higgs particle in SM couples appreciably only to the
singlet $t_R$ and the doublet $(t_L, b_L)$ fields, and of course
to the appropriate electroweak gauge bosons.
Motivated by this, we assume that the NP hidden in the scalar
sector is somehow able to discriminate among the
families (\eg\@ through some kind of a horizontal symmetry), 
and that it 
predominantly couples to the quarks of the third family.
Assuming in addition that NP is CP invariant and that its 
scale is large, we conclude that a reasonably general
description of  NP is in terms of a linear combination of all possible 
$dim=6$ $SU(3)\times SU(2) \times U(1)$ gauge invariant
and CP symmetric operators involving the scalar doublet of SM, 
together with the quarks of the third family and of course the gauge 
bosons\footnote{Note that
~appearance of the gauge bosons is an inevitable consequence
of the gauge principle and the presence of derivative
couplings.}
Thus, the first aim of the present work is to establish the
complete list of all possible such operators.
Conservatively, we have included in this list 
 also the operators containing the singlet $b_R$
field, in spite of the fact
that in SM it does not couple appreciably to the Higgs.
We should keep in mind though, that the
$b_R$ involving operators might be less likely to appear.
\par

The complete list of all  $dim=6$ $SU(3)\times
SU(2) \times U(1)$ gauge invariant  operators
involving all families of fermions
has been known for a long time \cite{Buchmuller}.
In establishing this list, the equations of motion have
traditionally been used in order to eliminate depended
operators. Within our philosophy though, in which the scalar 
doublet and the quarks of the third family have a very special
role in generating NP,  only those 
equations of motion which do not mix-in light quarks or leptons
are allowed to be used. Thus, in the approximation that we
neglect all fermion masses except the top, only the ~equations of
motion for the $t,~b$ quarks and the scalar fields are
to be used. Under such assumptions, the purely bosonic 
CP conserving operators have to a large extent 
already been classified in \cite{Hag-boson}; while 
two classes of CP conserving operators involving
the quarks of the third family and the gluon
have been given in \cite{topGRV1}.\par

Here we complete this list. Thus, we first add to the list of 
the purely
bosonic operators one more operator involving Higgs-gluon 
interactions and two purely gluonic ones.
Secondly, for the operators involving quarks of the third 
family, we complete the list  of  \cite{topGRV1},  
by adding to it also  the operators 
involving the covariant derivative of the
gauge  field strengths. Of course, after having identified the
NP operators, the gauge boson equations of motion may always
be used when doing calculations involving operators 
~containing the covariant derivative of the gauge  field strengths.
Because of this, the NP generated by the properties of the
third family, actually also induces NP couplings for the light
fermions as well,  
\cite{top-other, BLYoungH}.  \par

So in the first part of the
present work we give the complete list of all possible
such NP operators. It involves 14 purely bosonic operators and 34
operators containing quarks of the third family.
We then proceed to identify those that give
no tree level contributions to the present LEP1 and low energy
observables. 
These are the operators for which the present constraints 
are most mild, and which have therefore the
best chance to describe any possible kind of NP that may exist.
They are  9 bosonic and  25 quark operators.
Two of the bosonic operators involve only gluon fields, and
their study and compatibility with experiment necessitates a 
somewhat detail QCD treatment. Therefore we postpone their
consideration for the future, and in the present paper
concentrate on the 7 purely bosonic and the 25 quark 
operators.\par

The next step consists in studying the  
couplings associated to each NP operator 
at an energy scale smaller than the NP scale
$\lam$; ($s \lsim \lam^2 $). Remembering that the $dim=6$ nature of the
operator requires a dimensionful coupling constant
(\ie\@ involvement of a scale ), we 
remark that there are two possible ways to
introduce this effective Lagrangian: \par

(a) either from some theoretical prejudice
one knows the NP scale $\Lambda_{NP}$  and then one writes the 
interaction
\bq    
\L_{eff} = {f_i\over \Lambda^2_{NP}}\O_i \label{int1} \ \ , 
\eq\par

(b) or, 
if this is not the case, one chooses an arbitrary mass scale
$M$, (like \eg\@ $M_W$ or $m_t$) and writes
\bq 
\L_{eff} = {f_i\over M^2}\O_i   \label{int2} \ \ .
\eq
\noindent
If we follow the way "(a)", then an ambiguity often appears 
in the definition of $\lam$ and thus in the normalization of
$f_i$. So
it is difficult to accurately define the strength of the
interaction. If the underlying renormalizable dynamical theory which
induces the above effective NP Lagrangian were known, then the
"matching conditions" could be used to determine
$f_i(\lam^2)$ \cite{Georgi},
and the renormalization group equations would subsequently determine 
$f_i(s)$ at any lower scale $s \lsim \lam^2$.  
On the other hand, if the underlying renormalizable dynamical 
theory is not known, 
then we have no theoretical means to determine
$f_i(\lam^2)$. \par

Nevertheless, as proposed in Ref. \cite{unit1}, 
it is  possible to determine for any given value of $f_i$
in (\ref{int1}) or (\ref{int2}), the energy scale at which 
the interactions described by the corresponding operator 
$\O_i$, becomes "strong". The fact that these interactions have to become
strong at some energy, is an inevitable consequence of the $\O_i$
dimensionality being larger than four. 
Thus, for operators with $dim=6$ the tree level two-body
amplitudes generally grow like $s$, inevitably approaching the
unitarity limit at some energy scale. One can then define 
the NP scale $\Lambda_{NP}$ as the energy at which this happens.
This is a natural
definition, as at this energy one precisely expects that the residual
"effective"  description ceases to be valid. In other words,
for $s \sim \Lambda^2_{NP}$ or $s \lsim \Lambda^2_{NP}$, 
new particles, resonances, or  substructures,....,
typical of NP should appear. These are the phenomena that should
restore unitarity by changing the form of the interaction Lagrangian
$\L_{eff}$.\par

To achieve this procedure, we use the way "(b)" mentioned
above, and compute the unitarity constraint for each
interaction ${f_i\over M^2}\O_i$, using two-body amplitudes. 
One then gets
relations of the type
\bq  
{f_i\over M^2} \cdot {s\over C_i(s)} = 1    \ \ ,  \  \label{int3}
\eq
where $C_i(s)$ is a well-defined (generally) energy dependent 
coefficient. Defining then $\lam$ as  
\bq  
\Lambda_{NP}  \equiv \left [C_i(\lam^2) \cdot 
{M^2\over f_i}\right ]^{1/2} \label{int4} \ \ ,
\eq 
we rewrite the unitarity constraint in (\ref{int3}), as 
\bq  
f_i=  C_i(\lam^2) \cdot {M^2\over  \Lambda^2_{NP} } \ \ . 
\label{int5}  
\eq 
Note that there is no  ambiguity in the definition of $\lam$.
In other words, for any given $f_i$ and $\O_i$, $\lam $ is
exactly determined, at least at the level of using tree level 
amplitudes.\par

For the six purely
bosonic operators appearing in \cite{Hag-boson}, the unitarity 
relations have been established in 
\cite{unit1, Tsirigoti2, HZGRV} . Here we present
the results for the new bosonic operator $\O_{GG}$ inducing
anomalous Higgs-gluon interactions, and the  
set of the quark operators selected as explained above.\par

The contents of the paper is the following. In Sect.2 the complete
list of the gauge invariant operators is established. The unitarity
constraints are established in Sect.3. In Sect.4 the
implications of various renormalizable dynamical models on  
the appearance and strength of the various operators is 
presented, while the final discussion is given in Sect.5. \par

\section{The list of $dim=6$ gauge-invariant operators}  
The complete list of the CP conserving  purely bosonic
$dim=6$ and $SU(3)_c \times SU(2)\times U(1)$ gauge invariant
operators is given by
\bqa
\overline{\O}_{DW} & =& 2 ~ (D_{\mu} \overrightarrow W^{\mu
\rho}) (D^{\nu} \overrightarrow W_{\nu \rho})  \ \ \
  , \ \  \label{listDW}  \\[0.1cm]
\overline{\O}_{DG} & =& 2 ~ (D_{\mu} \overrightarrow G^{\mu
\rho}) (D^{\nu} \overrightarrow G_{\nu \rho})  \ \ \
  , \ \  \label{listDG}  \\[0.1cm]  
\O_G &= & {1\over3!}~  f_{ijk}~ G^{i\mu\nu}
  G^{j}_{\nu\lambda} G^{k\lambda}_{\ \ \ \mu} \ \ \
 ,  \ \ \   \label{listG} \\[0.1cm]  
\ol{\O}_{DB} & = & 2~(\partial_{\mu}B^{\mu \rho})(\partial^\nu B_{\nu
\rho}) \ \ \  , \ \   \label{listDB} \\[0.1cm] 
\O_{BW} & =& \frac{1}{2}~ \Phi^\dagger B_{\mu \nu}
\overrightarrow \tau \cdot \overrightarrow W^{\mu \nu} \Phi
\ \ \  , \ \  \label{listBW}  \\[0.1cm] 
\O_{\Phi 1} & =& (D_\mu \Phi^\dagger \Phi)( \Phi^\dagger
D^\mu \Phi) \ \ \  , \ \ \\[0.1cm]   \label{listPhi1} 
\O_{\Phi 2} & = & 4 ~ \partial_\mu (\Phi^\dagger \Phi)
\partial^\mu (\Phi^\dagger \Phi ) \ \ \  , \ \ \
  \label{listPhi2} \\[0.1cm] 
\O_{\Phi 3} & = & 8~ (\Phi^\dagger \Phi) ^3\ \ \  ,
\   \label{listPhi3} \\[0.1cm] 
\O_W &= & {1\over3!}\left( \overrightarrow{W}^{\ \ \nu}_\mu\times
  \overrightarrow{W}^{\ \ \lambda}_\nu \right) \cdot
  \overrightarrow{W}^{\ \ \mu}_\lambda \ \ \
 ,  \ \  \label{listW} \\[0.1cm]  
\O_{W\Phi} & = & i\, (D_\mu \Phi)^\dagger \overrightarrow \tau
\cdot \overrightarrow W^{\mu \nu} (D_\nu \Phi) \ \ \  , \ \
\label{listWPhi} \\[0.1cm]
\O_{B\Phi} & = & i\, (D_\mu \Phi)^\dagger B^{\mu \nu} (D_\nu
\Phi)\ \ \  , \ \ \label{listBPhi} \\[0.1cm]
\O_{WW} & = &  (\Phi^\dagger \Phi )\,    
\overrightarrow W^{\mu\nu} \cdot \overrightarrow W_{\mu\nu} \ \
\  ,  \ \ \label{listWW}\\[0.1cm]
\O_{BB} & = &  (\Phi^\dagger \Phi ) B^{\mu\nu} \
B_{\mu\nu} \ \ \  , \ \ \   \label{listBB} \\[0.1cm] 
\O_{GG} & = &  (\Phi^\dagger \Phi)\, 
\overrightarrow G^{\mu\nu} \cdot \overrightarrow G_{\mu\nu} \ \
\  .  \ \    \label{listGG} 
\eqa
Except of $\O_{GG}$,  $\ol{\O}_{DG}$ and $\O_G$,
these operators were first
enumerated in \cite{Hag-boson}, following the general
classifications in \cite{Buchmuller}. In connection with this we
also note that in \cite{Hag-boson}, instead of $\ol{\O}_{DW}$ 
and $\ol{\O}_{DB}$ the operators
\bq
\O_{DW}  =  ~ (D_{\mu} \overrightarrow W_{\nu
\rho}) (D^{\mu} \overrightarrow W^{\nu \rho})  \ \ \ , 
\ \ \ \O_{DB}  =  ~ (D_{\mu} B_{\nu \rho}) (D^{\mu} B^{\nu \rho}) 
   \ \    \label{listDWDB} 
\eq
are used, which satisfy 
\bq
\O_{DW}~=~ \ol{\O}_{DW} +12 g ~ \O_W \ \ \ \ , \ \  \O_{DB}~=~
\ol{\O}_{DB} \ . \   \label{listDWDB1} 
\eq
In the preceding ~formulae the usual definitions 
\bq 
\Phi=\left( \begin{array}{c}
      \phi^+ \\
{1\over\sqrt2}(v+H+i\phi^0) \end{array} \right) \ \ \ \ , \ \   
\label{listPhi} 	
\eq
\bq
D_{\mu}  =  (\partial_\mu + i~ g\prime\,Y B_\mu +
i~ \frac{g}{2} \overrightarrow \tau \cdot \overrightarrow W_\mu  +
i~ \frac{g_s}{2} \overrightarrow \lambda \cdot \overrightarrow G_\mu ) \ \
 , \  \label{listDmu} 
\eq
are used where $v\simeq 246 ~GeV$, $Y$ is the hypercharge of the
field on which the
covariant derivative acts, and $\overrightarrow \tau$ and  
$\overrightarrow \lambda$ are the isospin and colour
matrices applicable whenever $D_\mu$ acts on ~iso-doublet fields
and quarks respectively.\par

We next turn to the operators containing also quarks of the
third family which first appeared in \cite{topGRV1, Buchmuller}.
As in \cite{topGRV1}, we put in Class 1 those operators which involve
at least one $t_R$ field, but do not contain the covariant
derivative of the gauge boson field strength. Correspondingly in 
Class 2 we put the operators which contain neither $t_R$
nor any covariant derivative of the gauge boson 
field strengths. 
In both cases, the operators in each class are further divided 
into two groups containing four-quark and two-quark fields
respectively. Finally in Class 3 (which was not included in 
\cite{topGRV1}) we put the operators involving covariant
derivatives of gauge boson field strengths. It may be useful to
remark that all quark fields in 
(\ref{listqt}-\ref{listqG}) below, 
should be considered as weak eigen-fields for the
third family. Therefore,  the left-handed of them 
should eventually be mixed by the usual CKM matrix,
in order to give the mass eigen-state fields. 
We thus have:

\vspace{0.5cm}
{\bf Class 1.}\\
A1) \underline {Four-quark operators}
\bqa
\O_{qt} & = & (\bar q_L t_R)(\bar t_R q_L) \ \ \ , \ 
\label{listqt}\\[0.1cm] 
\O^{(8)}_{qt} & = & (\bar q_L \overrightarrow\lambda t_R)
(\bar t_R \overrightarrow\lambda q_L) \ \ \ ,\ \label{listqt8} \\[0.1cm]
\O_{tt} & = & {1\over2}\, (\bar t_R\gamma_{\mu} t_R)
(\bar t_R\gamma^{\mu} t_R) \ \ \ , \ \label{listtt}\\[0.1cm] 
\O_{tb} & = & (\bar t_R \gamma_{\mu} t_R)
(\bar b_R\gamma^{\mu} b_R) \ \ \ , \ \label{listtb}\\[0.1cm] 
\O^{(8)}_{tb} & = & (\bar t_R\gamma_{\mu}\overrightarrow\lambda t_R)
(\bar b_R\gamma^{\mu} \overrightarrow\lambda b_R) \ \ \ , 
\label{listtb8} \\[0.1cm]
\O_{qq} & = & (\bar t_R t_L)(\bar b_R b_L) +(\bar t_L t_R)(\bar
b_L b_R)\ \ \nonumber\\
\null & \null & - (\bar t_R b_L)(\bar b_R t_L) - (\bar b_L t_R)(\bar
t_L b_R) \ \ \ , \ \label{listqq}\\[0.1cm] 
\O^{(8)}_{qq} & = & (\bar t_R \overrightarrow\lambda t_L)
(\bar b_R\overrightarrow\lambda b_L)
+(\bar t_L \overrightarrow\lambda t_R)(\bar b_L
\overrightarrow\lambda  b_R)
\ \nonumber\\
\null & \null &
- (\bar t_R \overrightarrow\lambda b_L)
(\bar b_R \overrightarrow\lambda t_L)
- (\bar b_L \overrightarrow\lambda t_R)(\bar t_L
\overrightarrow\lambda   b_R)
\ \ \  . \label{listqq8} 
\eqa\\
B1) \underline {Two-quark operators.}
\bqa
\O_{t1} & = & (\Phi^{\dagger} \Phi)(\bar q_L t_R\widetilde\Phi +\bar t_R
\widetilde \Phi^{\dagger} q_L) \ \ \ ,\ \label{listt1} \\[0.1cm]
\O_{t2} & = & i\,\left [ \Phi^{\dagger} (D_{\mu} \Phi)- (D_{\mu}
\Phi^{\dagger})  \Phi \right ]
(\bar t_R\gamma^{\mu} t_R) \ \ \ ,\label{listt2} \\[0.1cm]
\O_{t3} & = & i\,( \widetilde \Phi^{\dagger} D_{\mu} \Phi)
(\bar t_R\gamma^{\mu} b_R)-i\, (D_{\mu} \Phi^{\dagger}  \widetilde\Phi)
(\bar b_R\gamma^{\mu} t_R) \ \ \ ,\label{listt3} \\[0.1cm]
\O_{D t} &= & (\bar q_L D_{\mu} t_R)D^{\mu} \widetilde \Phi +
D^{\mu}\widetilde \Phi^{\dagger}
(\ol{D_{\mu}t_R}~ q_L) \ \ \ , \label{listDt}\\[0.1cm] 
\O_{tW\Phi} & = & (\bar q_L \sigma^{\mu\nu}\overrightarrow \tau
t_R) \widetilde \Phi \cdot
\overrightarrow W_{\mu\nu} + \widetilde \Phi^{\dagger}
(\bar t_R \sigma^{\mu\nu}
\overrightarrow \tau q_L) \cdot \overrightarrow W_{\mu\nu}\ \ \
,\label{listtWPhi}\\[0.1cm] 
\O_{tB\Phi}& = &(\bar q_L \sigma^{\mu\nu} t_R)\widetilde \Phi
B_{\mu\nu} +\widetilde \Phi^{\dagger}(\bar t_R \sigma^{\mu\nu}
 q_L) B_{\mu\nu} \ \ \ ,\label{listtBPhi}\\[0.1cm]
\O_{tG\Phi} & = & \left [ (\bar q_L \sigma^{\mu\nu} \lambda^a t_R)
\widetilde \Phi
 + \widetilde \Phi^{\dagger}(\bar t_R \sigma^{\mu\nu}
\lambda^a q_L)\right ] G_{\mu\nu}^a  \ \ \ . \ \label{listtGPhi}
\eqa\\

\vspace{0.5cm}
{\bf Class 2.}\\
A2) \underline{Four quark operators}
\bqa
\O^{(1,1)}_{qq} & = &{1\over2}\, (\bar q_L\gamma_{\mu} q_L)
(\bar q_L\gamma^{\mu} q_L) \ \ \ , \label{listqq11}\\[0.1cm]
\O^{(1,3)}_{qq} & = & {1\over2}\,(\bar
q_L\gamma_{\mu}\overrightarrow\tau q_L) \cdot
(\bar q_L\gamma^{\mu} \overrightarrow\tau q_L) \ \ \ ,
\label{listqq13}\\[0.1cm]
\O_{bb} & = & {1\over2}\,(\bar b_R\gamma_{\mu} b_R)
(\bar b_R\gamma^{\mu} b_R) \ \ \ ,\\[0.1cm]\label{listbb}
\O_{qb} & = & (\bar q_L b_R)(\bar b_R q_L) \ \ \ ,
\label{listqb} \\[0.1cm]
\O^{(8)}_{qb} &= & (\bar q_L\overrightarrow\lambda b_R)\cdot
(\bar b_R\overrightarrow\lambda q_L) \ \ \ . \ \label{listqb8}
\eqa\\
B2) \underline {Two-quark operators.}
\bqa
\O^{(1)}_{\Phi q} &= & i\,(\Phi^{\dagger}  D_{\mu} \Phi)
(\bar q_L\gamma^{\mu} q_L) -i\,(D_{\mu} \Phi^{\dagger}  \Phi)
(\bar q_L\gamma^{\mu} q_L) \ \ \ ,\label{listPhiq1}\\[0.1cm]
\O^{(3)}_{\Phi q} &=& i\,\left [( \Phi^{\dagger}
\overrightarrow \tau D_{\mu} \Phi)
-( D_{\mu} \Phi^{\dagger} \overrightarrow \tau \Phi)\right ]
\cdot (\bar q_L\gamma^{\mu}\overrightarrow\tau q_L) \ \ \ ,
\label{listPhiq3}\\[0.1cm] 
\O_{\Phi b}& = &i\,\left [( \Phi^{\dagger}  D_{\mu} \Phi)
-( D_{\mu} \Phi^{\dagger}\Phi)\right ]
(\bar b_R\gamma^{\mu} b_R) \ \ \ ,\label{listPhib}\\[0.1cm]
\O_{D b} &=& (\bar q_L D_{\mu} b_R)D^{\mu}\Phi + D^{\mu}\Phi^{\dagger}
(\ol{D_{\mu}b_R} q_L) \ \ \ ,\label{listDb}\\[0.1cm]
\O_{bW\Phi} &=& (\bar q_L \sigma^{\mu\nu}\overrightarrow \tau
b_R)\Phi \cdot
\overrightarrow W_{\mu\nu} + \Phi^{\dagger}(\bar b_R \sigma^{\mu\nu}
\overrightarrow \tau q_L)\cdot \overrightarrow W_{\mu\nu}
\ \ \ ,\label{listbWPhi}\\[0.1cm]
\O_{bB\Phi} &= & (\bar q_L \sigma^{\mu\nu} b_R)\Phi
B_{\mu\nu} + \Phi^{\dagger}(\bar b_R \sigma^{\mu\nu}
 q_L) B_{\mu\nu} \ \ \ ,\label{listbBPhi}\\[0.1cm]
\O_{bG\Phi} &=& (\bar q_L \sigma^{\mu\nu}\lambda^a b_R)\Phi
G_{\mu\nu}^a + \Phi^{\dagger}(\bar b_R \sigma^{\mu\nu}
\lambda^a q_L)G_{\mu\nu}^a \ \ \ ,\label{listbGPhi}\\[0.1cm]
\O_{b1} & = & (\Phi^{\dagger} \Phi)(\bar q_L b_R \Phi +\bar b_R
 \Phi^{\dagger} q_L) \ \ \ .\ \ \label{listb1}
\eqa\\

\vspace{0.5cm}
{\bf Class 3.}
\bqa
\O_{qB} & = & \bar q_L \gamma^\mu q_L (\partial^\nu B_{\mu\nu}) \ \
\ , \label{listqB}\\[0.1cm]
\O_{qW}& =& \frac{1}{2} \left (\bar q_L \gamma_{\mu}
\overrightarrow\tau q_L \right ) \cdot
(D_\nu \overrightarrow W^{\mu\nu}) \ \ \ , \label{listqW}\\[0.1cm]
\O_{bB} & = & \bar b_R \gamma^\mu b_R (\partial^\nu B_{\mu\nu}) \ \
\ , \label{listbB}\\[0.1cm]
\O_{tB} & = & \bar t_R \gamma^\mu t_R (\partial^\nu B_{\mu\nu}) \ \
\ , \label{listtB}\\[0.1cm]
\O_{tG}& =& \frac{1}{2} \left (\bar t_R\gamma_{\mu}
\overrightarrow\lambda t_R \right ) \cdot
(D_\nu \overrightarrow G^{\mu\nu}) \ \ \ , \label{listtG}\\[0.1cm]
\O_{bG}& =& \frac{1}{2} \left (\bar b_R\gamma_{\mu}
\overrightarrow\lambda b_R \right ) \cdot
(D_\nu \overrightarrow G^{\mu\nu}) \ \ \ , \label{listbG}\\[0.1cm]
\O_{qG}& =& \frac{1}{2} \left (\bar q_L \gamma_{\mu}
\overrightarrow\lambda q_L \right ) \cdot
(D_\nu \overrightarrow G^{\mu\nu}) \ \ \ . \ \label{listqG}
\eqa
where $\lambda^a$ are the eight usual colour matrices.\par

Concerning the above list, a few remarks must be made.
As mentioned already,  we have used the
equations of motion for the quark and scalar fields, but not for
the gauge bosons, since the later mix-in light fermions. 
This attitude leads to including in the list 
also the operators  $\O_{qB}$, $\O_{qW}$, $\O_{bB}$, $\O_{tB}$, 
$\O_{tG}$, $\O_{bG}$ and $\O_{qG}$, collected in Class 3 and defined
through  (\ref{listqB}-\ref{listqG}). 
Eqs. (\ref{listqB}-\ref{listqG}) constitute \underline{one possible
definition} for these operators though.  Another possibility is 
to substitute  for them in (\ref{listqB}-\ref{listqG})  the gauge 
boson equations of motion
\bqa
D_\mu \overrightarrow G^{\mu\nu} & = & g_s 
\overrightarrow J^\nu_{(3)}\ \ ,\label{listDmuG} \\[0.1cm]
D_\mu \overrightarrow W^{\mu\nu} & = & g 
\overrightarrow J^\nu_{(2)}~ -~ i~\frac{g}{2} ~[D^\nu \Phi^\dagger
\overrightarrow \tau \Phi ~ -~ \Phi^\dagger
\overrightarrow \tau D^\nu \Phi ]\ \ , \label{listDmuW}\\[0.1cm] 
\partial_\mu  B^{\mu\nu} & = & g\prime 
J^\nu_{(1)}~ -~ i~\frac{g\prime}{2}~ [D^\nu \Phi^\dagger \Phi ~ - 
~\Phi^\dagger  D^\nu \Phi ]\ \ , \ \label{listDmuB}
\eqa
where $\overrightarrow J^\nu_{(3)}$, $\overrightarrow
J^\nu_{(2)}$ and
$J^\nu_{(1)}$ are the $SU(3)$, $SU(2)$ and hypercharge fermionic
currents respectively. Such a substitution provides
\underline{another possible definition} for these operators
which is in fact more convenient for higher order 
calculations \cite{Georgi}. At tree level and to linear order
in the NP couplings both definitions give identical results in
Feynman diagram calculations. Differences start appearing 
at higher order, for which the definition using directly 
 (\ref{listqB}-\ref{listqG})  
implies a different (and more involved) structure of 
counter-terms\footnote{These remarks are of course also valid for
$\ol{\O}_{DW}$, $\ol{\O}_{DB}$ and $\ol{\O}_{DG}$ given in
(\ref{listDW},\ref{listDB},\ref{listDG}) 
respectively.} \cite{Georgi}. 
We come back to this below.\par

Compared 
to \cite{topGRV1} we should also mentioned that we have
dropped the operators 
\bqa
\O^{(8,1)}_{qq} & = & {1\over2}\,(\bar
q_L\gamma_{\mu}\overrightarrow\lambda q_L).
(\bar q_L\gamma^{\mu} \overrightarrow\lambda q_L) \ \ \
,\label{listqq81}\\[0.1cm]
\O^{(8,3)}_{qq} & = & {1\over2}\,(\bar q_L\gamma_{\mu}\lambda^a\tau^j q_L)
(\bar q_L\gamma^{\mu}\lambda^a\tau^j q_L) \ \ \ , \ \label{listqq83}
\eqa
since they are related to $\O^{(1,1)}_{qq}$, $\O^{(1,3)}_{qq}$
in  (\ref{listqq11}, \ref{listqq13}) 
through the Fierz identities
\bqa
\O^{(8,1)}_{qq} & = & \O^{(1,3)}_{qq} +\frac{1}{3}~
\O^{(1,1)}_{qq}  \ \ \, \label{listqq813}\\[0.1cm]
\O^{(8,3)}_{qq} & = &
3\O^{(1,1)}_{qq}-\frac{5}{3}~ \O^{(1,3)}_{qq}  \ \ \ . \ 
\label{listqq831}
\eqa
Finally we have also added for completeness the 2nd Class operator 
$\O_{b1}$ (analogous to $\O_{t1}$) which is $\Phi^\dagger \Phi$ 
times the standard model Yukawa mass term for the $b$-quark.
We should keep in mind though, that the $b_R$ involving NP
operators are on a somewhat weaker basis, 
since SM suggests that $b_R$ couples very weakly to the Higgs
field, which is assumed to be the source of NP. \par

In the framework explained so far, NP is described in terms of 14
purely bosonic CP conserving $dim=6$ operators, and 34 operators
involving quarks of the third family. To proceed further we
reduce the number of operators to be studied by excluding those
contributing at tree level to LEP1 and lower energy
observables. Thus, we 
remark that $\ol{\O}_{DW}$ contributes at tree level to 
$\epsilon_{1,2,3}$, 
$\ol{\O}_{DB}$ to $\epsilon_{1,3}$, 
$\ol{\O}_{BW}$ to $\epsilon_3$ and $\O_{\Phi1}$ to $\epsilon_1$
\cite{Altarelli, Hag-boson, DeR}.
Moreover, the 2nd and 3rd Class operators $\O^{(1)}_{\Phi q}$,
$\O^{(3)}_{\Phi q}$, $\O_{\Phi b}$, $\O_{Db}$, $\O_{bW\Phi}$,
$\O_{bB\Phi}$, $\O_{qB}$, $\O_{qW}$ and $\O_{bB}$ give tree
level contributions to $Zb\bar b$ and they are thus also very
strongly constrained \cite{topGRV1}; so far as we consider the action
of one operator at a time. In addition, we also exclude from any
further consideration the operator $\O_{\Phi3}$ 
(\ref{listPhi3}), since it gives no contribution to LEP1
physics and its experimental study in the future Colliders looks 
almost impossible.  
Therefore the operators which are not already very strongly 
constrained by existing measurements are the 9 bosonic
ones in (\ref{listDG}, \ref{listG}, \ref{listPhi2}, 
\ref{listW} -\ref{listGG}) 
and the 25 quark operators in (\ref{listqt}-\ref{listqb8},
\ref{listbGPhi} , \ref{listb1}, \ref{listtB}-\ref{listqG}).
In the following we study the unitarity constraints on the 
couplings of all these operators except for the purely gluonic ones 
$\ol{\O}_{DG}$ and $\O_G$, which are left for a  
future work as this deserves a special treatment of QCD effects 
(see the final discussion).\par   

\section{Unitarity constraints}
Following the procedure "(b)" explained in the Introduction, 
we define the New Physics couplings through the effective
lagrangian 
\bq
\L_{eff} = \L_t + \L_{bos} \ \ , \  \label{listLLL}
\eq
where  
the contribution from the 25 ($i=1 ... 25$) "quark" operators 
is written as
\bq
\L_t =  \sum_i { f_i \over m^2_t}\,\O_i \ \ \ . \ \label{listLt}
\eq
As in (\ref{int2}), the $\mt$ in the denominators in 
(\ref{listLt}) are simply normalization factors. Correspondingly,
 the contribution from the 7 purely bosonic operators  is 
written as\footnote{The operators $\O_{WW}$
$\O_{BB}$ are analogous to the $\O_{UW}$ and $\O_{UB}$
introduced in
\cite{NP-GLR, unit1, Tsirigoti2, HZGRV}, but the definition of 
their couplings is exactly the same.}  
\bqa
{\L}_{bos} & = & \lambda_W {g\over M^2_W}\O_W+f_W{g\over2M^2_W}
\O_{W\Phi}+f_B{g'\over2M^2_W}  \O_{B\Phi}+ \nonumber \\
 & &\frac{d}{v^2} \ {\cal O}_{WW}+\frac{d_B}{v^2}
\ {\cal O}_{BB}+ \frac{f_{\Phi 2}}{v^2}
{\cal O}_{\Phi 2}~+~ \frac{d_G}{v^2} \ {\cal O}_{GG}\ \  \ \ .\ 
\label{listLbos}
\eqa\par

The unitarity constraints for the first 6 purely bosonic
operators has been done in \cite{unit1, Tsirigoti2, HZGRV}.
The results are
\bq
|\lambda_W| \simeq 19~{M^2_W \over \lam^2} \ \ \ \ \ , \ \ \ \ \ \ \
|f_B| \simeq 98~{M^2_W\over\lam^2} \ \ \ \ \ , \ \ \ \ \
 \ |f_W| \simeq 31~{M^2_W \over\lam^2} \ \ \ \ ,\
\eq
\begin{eqnarray}
d & \simeq & \frac{104.5~\left ({\frac{M_W}{\Lambda_{NP}}}
\right )^2} {1+6.5 \left ({\frac{M_W}{\Lambda_{NP}}}\right )} \ \
\mbox{ for } d>0 \ , \nonumber \\
d & \simeq & -~ \frac{104.5~\left ({\frac{M_W}{\Lambda_{NP}}}
\right )^2} {1- 4 \left ({\frac{M_W}{\Lambda_{NP}}}\right )} \ \
\mbox{ for }\  d<0 \ , \  \\
d_B & \simeq & \frac{195.8 ~\left ({\frac{M_W}{\Lambda_{NP}}}
\right )^2} {1+200 \left  ({\frac{M_W}{\Lambda_{NP}}}
\right )^2}\ \
\mbox{ for } d_B>0 \ , \nonumber \\
d_B & \simeq & -~ \frac{195.8 ~\left ({\frac{M_W}{\Lambda_{NP}}}
\right )^2} {1 +50 \left  ({\frac{M_W}{\Lambda_{NP}}}
\right )^2}\ \
\mbox{ for }\  d_B<0 \ , \  
\eqa
while for $\O_{\Phi 2}$ we refer to \cite{HZGRV}. \par

For the bosonic operator $\O_{GG}$, (which induces
a custodial $SU(2)_c$ invariant Higgs-gluon coupling), the strongest
unitarity constraint arises from the colour singlet $J=0$
channels $|gg++>$, $|gg-->$ and $|HH>$. Diagonalizing the
transition matrix we get for it
\bq
|d_G| \simeq \frac{4\pi}{\sqrt{1+\frac{60 \pi v^2}{\lam^2}}}~
 \left (\frac{v}{\lam}\right )^2
\simeq \frac{119 }{\sqrt{1+ \frac{1782\mwd}{\lam^2}}}~
 \left (\frac{\mw}{\lam}\right )^2 \ .
\eq\par

We next turn to the selected 25 operators, involving quarks 
of the third
family. As said above, 14 of them belong to the first class,
7 to the second class and 4 to the third one. We start from the
14 1st Class operators. We first give the unitarity 
relations and subsequently we discuss them. These are 
\bqa
|f_{qt}| & \simeq & {16\pi\over3}~
\left ({m^2_t\over \Lambda^2_{NP}}\right ) \ \ ,
\label{fqt}\\[0.1cm]
|f^{(8)}_{qt}| & \simeq & {9\pi\over\sqrt2}~
\left ({m^2_t\over \Lambda^2_{NP}}\right ) \ \ , 
\label{fqt8} \\[0.1cm]
|f_{tt}| & \simeq & 6\pi~ \left ({m^2_t\over \Lambda^2_{NP}}
\right ) \ \ ,\\[0.1cm]
|f_{tb}| & \simeq & 8\pi ~ \left ({m^2_t\over \Lambda^2_{NP}}
\right ) \ \ ,\\[0.1cm]
|f^{(8)}_{tb}| & \simeq & {9\pi\over 2}
\left ({m^2_t\over \Lambda^2_{NP}} \right )\ \ , \\[0.1cm] 
|f_{qq}| & \simeq & {32\pi\over7}~
\left ({m^2_t\over \Lambda^2_{NP}}\right ) \ \ , \\[0.1cm]
 |f^{(8)}_{qq}| & \simeq & 6\pi ~ \left ({m^2_t\over 
\Lambda^2_{NP}} \right ) \ \ , \\[0.1cm]
|f_{t1}| & \simeq & {16\pi\over3\sqrt2}~
\left ({m^2_t\over v\Lambda_{NP}} \right ) \ \ , 
\label{ft1} \\[0.1cm]
|f_{t2}| & \simeq & 8\pi\sqrt3 ~
\left ({m^2_t\over \Lambda^2_{NP}}\right ) \ \ , \\[0.1cm]
|f_{t3}| & \simeq & 8\pi\sqrt6 ~
\left ({m^2_t\over \Lambda^2_{NP}} \right )\ \ , \\[0.1cm]
|f_{Dt}| & \simeq & 10.4  
\left ({m^2_t\over \Lambda^2_{NP}} \right ) \ \ \mbox{for }
f_{Dt} >0  \nonumber \\[0.1cm]
|f_{Dt}| & \simeq & -~6.4
\left ({m^2_t\over \Lambda^2_{NP}} \right ) \ \ \mbox{for }
f_{Dt} <0 , \\[0.1cm]
|f_{tW\Phi}| & \simeq & {61.6\over\sqrt{1+645\, {m^2_t\over 
\Lambda^2_{NP}}}}~
\left ({m^2_t\over \Lambda^2_{NP}} \right) \ \ , \\[0.1cm]
|f_{tB\Phi}| & \simeq &  {61.6\over\sqrt{1+645\, {m^2_t\over 
\Lambda^2_{NP}}}}~
\left ({m^2_t\over \Lambda^2_{NP}} \right) \ \ , \\[0.1cm]
|f_{tG\Phi}| & \simeq &
{m^2_t \sqrt{\pi}\over v \lam \sqrt{1+\frac{2}{3}\alpha_s}} \ \
\ \mbox{for } \ \lam\lsim 10TeV \ \nonumber\\[0.1cm]
|f_{tG\Phi}| & \simeq &
{75 (m_t/\lam)^2 \over \sqrt{1+591~(m_t/\lam)^2}} \ \
\ \mbox{for } \ \lam\gsim 10TeV \ . \
\eqa\par

These results arise as follows:\par

$\O_{qt}$: The dominant unitarity constraint arises from the $J=0$
transition amplitude affecting the colour singlet channel
$|t\bar t ++>$ or $|t\bar b ++>$.\par

$\O^{(8)}_{qt}$: The dominant constraint comes from the $J=1$
transition matrix
affecting the colour singlet channels $|t\bar t +->$, $|t\bar t
-+>$, $|b\bar b -+>$.\par

$\O_{tt}$: From the $J=1$ transition amplitude affecting the colour 
singlet $|t\bar t +->$.\par

$\O_{tb}$: From the $J=1$ transition matrix affecting the colour
singlet $|t\bar t +->$, $|b\bar b +->$ channels.\par

$\O^{(8)}_{tb}$: From the $J=1$ transition amplitude affecting 
the colour singlet $|t\bar b +->$ channel.\par   
 
$\O_{qq}$: From the $J=0$ 
transition matrix affecting any of the colour
singlet sets of channels $(|t\bar t ++> ~, ~|b\bar b -->)$,
$ (|t\bar t --> ~,~ |b\bar b ++>)$ or 
$(|b\bar t ++> ~,~ |b\bar t -->)$.\par

$\O^{(8)}_{qq}$: From the $J=0$ transition matrix 
affecting any of the colour
singlet sets of channels $(|t\bar t ++> ~, ~|b\bar b -->)$,
or  $(|b\bar t ++> ~,~ |b\bar t -->)$.\par

$\O_{t1}$: This operator is of the form $\Phi^\dagger \Phi$ 
times the standard Yukawa top mass term. Although $\O_{t1}$ is 
formally a $dim=6$ operator, it actually behaves as a
lower-dimension one when restricting to two-particle channels. 
When this is nevertheless done, the dominant 
constraint comes from the $J=0$
transition ~matrix affecting the colour singlet channels 
$|t\bar t ++>$, $|t\bar t -->$, $|HH>$, 
$|W^+W^-~ LL>$, $|ZZ~LL>$. As seen from (\ref{ft1}), this leads to a
unitarity constraint in which the anomalous coupling
does not behave like $\sim 1/\lam^2$ for large $\lam$
\cite{BLYoungOt1}. 
A constraint of the form $\sim 1/\lam^2$ could only
be obtained by considering transitions affecting channels
containing more particles. The unitarity properties of $\O_{t1}$
are not further investigated here, since this operator 
gives no contribution
either to LEP1 observables \cite{topGRV1}, or to observables in
$e^-e^+ \to t \bar t$ and $t \to b W$ \cite{eettGKR}. \par

$\O_{t2}$: The dominant constraint arises from the $J=1$ 
transition matrix affecting the colour singlet channels 
$|t \bar t +->$, $|ZH~L>$, $|W^+W^-~ LL>$.\par

$\O_{t3}$: From the $J=1$ 
transition matrix affecting the colour singlet channels 
$|b \bar t +->$, $|W^- H~L>$, $|W^- Z~ LL>$. The transition matrix
elements may be obtained from the corresponding ones for 
$\O_{t2}$ by dividing them by $-\sqrt 2$.\par

$\O_{Dt}$: From the $J=0$
transition ~amplitude affecting the colour singlet channel
$|t\bar t ++>$ or $|t\bar t -->$.\par

$\O_{tW\Phi}$, $\O_{tB\Phi}$: From 
the $J=1$ transition matrices affecting the colour singlet channels
$|t\bar t ++>$ and $|t\bar t -->$.\par 

$\O_{tG\Phi}$: For $\lam \lsim 10TeV$, the dominant unitarity
constraint arises from the $J=0$ transition matrix affecting the
colour singlet channels  $|t\bar t ++>$,  $|t\bar t -->$,
$|gg ++>$ and $|gg -->$. For $\lam \gsim 10TeV$, the $J=1$
transition matrix affecting the colour-octet channels
$|t\bar t ++>$,  $|t\bar t -->$, $|gH +>$ and $|gH ->$,
dominates.\par

At this point, we have finished with the first Class operators
and we turn to the 2nd Class ones given in
(\ref{listqq11}-\ref{listqb8}, \ref{listbGPhi},
\ref{listb1}) which imply 
\bqa
|f^{(1,1)}_{qq}| & \simeq & {24 \pi\over 7}~
\left ({m^2_t\over \Lambda^2_{NP}}\right ) \ \ ,\\[0.1cm]
|f^{(1,3)}_{qq}| & \simeq & {24 \pi\over 5}~
\left ({m^2_t\over \Lambda^2_{NP}}\right ) \ \ , \\[0.1cm]
|f_{bb}| & \simeq & 6\pi~ \left ({m^2_t\over \Lambda^2_{NP}}
\right ) \ \ ,\\[0.1cm]
|f_{qb}| & \simeq & \frac{16 \pi}{3} ~ \left ({m^2_t\over \Lambda^2_{NP}}
\right ) \ \ ,\\[0.1cm]
|f^{(8)}_{qb}| & \simeq & {9\pi\over \sqrt 2}
\left ({m^2_t\over \Lambda^2_{NP}} \right )\ \ , \\[0.1cm] 
|f_{bG\Phi}| & \simeq &
{m^2_t \sqrt{\pi}\over v \lam \sqrt{1+\frac{2}{3}\alpha_s}} \ \
\ \mbox{for } \ \lam\lsim 10TeV \ \nonumber\\[0.1cm]
|f_{bG\Phi}| & \simeq &
{75 (m_t/\lam)^2 \over \sqrt{1+591~(m_t/\lam)^2}} \ \
\ \mbox{for } \ \lam\gsim 10TeV \ , \\[0.1cm]
|f_{b1}| & \simeq & {16\pi\over3\sqrt2}~
\left ({m^2_t\over v\Lambda_{NP}} \right ) \ \ . \
\eqa \par

These results arise as follows:\par

$\O^{(1,1)}_{qq}$: From the $J=1$ transition matrix affecting 
the colour singlet channels $|t\bar t -+>$, $|b\bar b -+>$.\par

$\O^{(1,3)}_{qq}$: From the $J=1$ transition matrix affecting 
the colour singlet channels $|t\bar t -+>$, $|b\bar b -+>$,
or the channel $|t\bar b -+>$.\par
 
$\O_{bb}$: From the $J=1$ transition amplitude affecting 
the colour singlet $|b \bar b +->$ channel; 
(similar to $\O_{tt}$).\par

$\O_{qb}$: From the $J=0$ transition ~amplitude affecting the
colour singlet channel $|b\bar b ++>$ or $|b \bar t ++>$;
(similar to $\O_{qt}$).\par

$\O^{(8)}_{qb}$: From the $J=1$ transition matrix affecting 
the colour singlet channels $|b \bar b +->$, $|b \bar b -+>$, 
$|t \bar t -+>$; (similar to $\O^{(8)}_{(qt)}$).\par

$\O_{bG\Phi}$: It is similar to $\O_{tG\Phi}$. 
For $\lam \lsim 10TeV$, the dominant constraint comes from the 
$J=0$ transition matrix affecting the
colour singlet channels  $|b \bar b ++>$,  $|b \bar b -->$,
$|gg ++>$ and $|gg -->$. For $\lam \gsim 10TeV$, the $J=1$
transition matrix affecting the colour-octet channels
$|b\bar b ++>$,  $|b\bar b -->$, $|gH +>$ and $|gH ->$,
dominates.\par

$\O_{b1}$: This operator is of the form $\Phi^\dagger \Phi$ 
times the standard Yukawa $b$-quark mass term.  $\O_{b1}$ 
is similar to $\O_{t1}$ and the same remarks apply. 
The dominant constraint, generated from
two-particle channels, comes from the colour singlet states 
$|b\bar b ++>$, $|b\bar b -->$, $|HH>$, 
$|W^+W^-~ LL>$, $|ZZ~LL>$.\par

The last set of unitarity constraints concerns the four 
3rd Class 
operators $\O_{tB}$, $\O_{tG}$, $\O_{bG}$, $\O_{qG}$ in
(\ref{listtB}-\ref{listqG}).
As already stated, we could define these operators either
directly by eq~(\ref{listtB}-\ref{listqG}), 
or alternatively by substituting in 
them the covariant derivative of the gauge boson field ~strengths
appearing in (\ref{listDmuG}-\ref{listDmuB}). Since these two 
definitions are only
identical to linear order in the anomalous couplings and in
general induce a different structure of counter terms at 
loop calculations, they can in principle also imply different 
unitarity relations. As we will see below, this
actually happens only for the $\O_{tB}$ case though. To explain these, we
start from the second definition (utilizing the gauge boson
equations of motion), which leads to the unitarity constraints
\bqa
|f_{tB}| & \simeq & 25 ~
\left ({m^2_t\over \Lambda^2_{NP}}\right ) \ \ ,
\label{ftB} \\[0.1cm]
|f_{tG}| & \simeq & \frac{6 \pi}{g_s}~
\left ({m^2_t\over \Lambda^2_{NP}}\right ) \ \ , 
\label{ftG}\\[0.1cm]
|f_{bG}| & \simeq & \frac{6 \pi}{g_s}~
\left ({m^2_t\over \Lambda^2_{NP}}\right ) \ \ , 
\label{fbG}\\[0.1cm]
|f_{qG}| & \simeq & \frac{6 \pi}{g_s}~
\left ({m^2_t\over \Lambda^2_{NP}}\right ) \ \ . \
\label{fqG}
\eqa\par

~Concerning (\ref{ftB}-\ref{fqG}), the following comments 
are in order:\par

$\O_{tB}$: When the $B_\mu$ equation of motion (see (\ref{listDmuB}))
is substituted in (\ref{listtB}), then the dominant
constraint arises from the $J=1$ transition matrix affecting the
colour singlet channels $|f \bar f +->$, $|f  \bar f -+>$, where 
$f$ is any fermion (quark or lepton), and $|ZH~L>$. All
transition matrix elements obtained this way, depend linearly on
$f_{tB}$ and lead to the constrain given in (\ref{ftB}). 
If instead (\ref{listDmuB}) is not
used, then there is an additional $f^2_{tB}$-contribution to the
amplitude $<t\bar t +-|T^{J=1}|t\bar t +->$, which changes the
result of (\ref{ftB}) as 
\bqa
f_{tB} & \simeq & 5.5 ~
\left ({m^2_t\over \Lambda^2_{NP}}\right ) \ \ \mbox{ for }
\ f_{tB}>0 \ , \nonumber \\[0.1cm]\label{listftB1}
f_{tB} & \simeq & - 4.8 ~
\left ({m^2_t\over \Lambda^2_{NP}}\right ) \ \  \mbox{ for }
f_{tB}<0 \ . \  \label{listftB2}
\eqa
At this point we would not like to enter arguments on
whether we should use  (\ref{listDmuB}) or not. 
Instead, we simply take the 
weaker constraint  (\ref{ftB}) as an indication on what could be 
the largest  allowed value of $f_{tB}$ for any given $\lam$.\par

$\O_{tG}$: The dominant unitarity constraint arises from the $J=0$
transition ~amplitude affecting the colour singlet channel
$|t\bar t ++>$. \par

$\O_{bG}$: Similar to $\O_{tG}$. The dominant unitarity constraint 
arises from the $J=0$ transition ~amplitude affecting the colour
singlet channel $| b\bar b ++>$.\par

$\O_{qG}$: Similar to the two previous ones. The dominant unitarity 
constraint arises from the $J=0$ transition ~amplitude affecting 
the colour singlet channel $|t\bar t -->$ or $|b\bar b -->$.\par

The results on the last three operators $\O_{tG}$, $\O_{bG}$,
$\O_{qG}$ do not depend on whether we define them through 
(\ref{listtB}-\ref{listqG}) directly, or after having
substituted in them the gauge
boson equations of motion appearing in
(\ref{listDmuG}-\ref{listDmuB}).\par

\section{Dynamical Scenarios} 

The aim of the present section is to investigate  
which of the above operators will be generated in a wide class 
of renormalizable dynamical models. Considerations of this type 
have already been presented in \cite{Tsirigoti1, Tsirigoti2}
for dynamical models inducing only purely bosonic operators 
and in \cite{Arzt} where fermionic operators are also 
generated, but no special role
is assigned to the quarks of the third family.\par

Here we concentrate on the NP operators observable through 
new anomalous couplings at the $0.5-1~TeV$ scale. We assume that
any possible new gauge bosons that might exist have much 
heavier masses. Therefore the relevant gauge group in the
$0.5-1~TeV$ region is simply
$SU(3)\times SU(2) \times U(1)$. Based on this, we  consider 
the most general renormalizable dynamical models containing SM
and involving 
in addition any number of scalars or fermions which are singlets
or doublets under weak isospin, and singlets or triplets under
colour. We always assume the new particles to get their masses
independently of the spontaneous breaking of the electroweak gauge
symmetry, which then leads to the natural expectation that
these masses should be sufficiently heavier than $v \simeq 246~
GeV$. For the new bosons this is always possible. In order to be
possible for the new fermions also we assume them to be
(in general) left-right symmetric, which also guarantees that
no anomalies are introduced.
Finally we note that the new scalars and fermions couple to 
the gauge bosons according to the gauge principle, while their 
couplings among themselves and the Higgs is determined by new,
often unknown, Yukawa-type couplings.  \par

When integrating out the aforementioned heavy fields, we find
that \cite{Papadamou-dyn, Tsirigoti1, Tsirigoti2}
\begin{itemize}
\item
the purely bosonic operators $\O_{W\Phi}$, 
$\O_{B\Phi}$ 
\item
the four-quark operator $\O_{qq}^{(8)}$, and
\item
the two-quark oprators $\O_{Dt}$, $\O_{Db}$, 
$\O_{bW\Phi}$, $\O_{bB\Phi}$ and
$\O_{bG\Phi}$ 
\end{itemize}
are never generated in any such model up to the 1-loop
order.\par 

For the rest of the operators listed above, we find that they
are generated in at least some of the models. In many cases
though, the generation or not of certain operators depends 
directly on the
quantum numbers of the heavy particles we have integrated out.
Thus, $\O_W$ is generated only when one of these heavy particles
carries isospin, while the appearance of
$\O_G$ is possible only when it carries colour. Corresponding
remarks apply also for the generation of $\O_{WW}$ and $\O_{GG}$
respectively, with the additional caveat that the responsible
heavy particle which has been integrated out must be scalar.\par

Concerning the couplings of the generated operators, we find
that those involving only gauge bosons like \eg\@
$\O_W$ or $\O_G$, have strengths
proportional to $g^n/\lam^2$, with $g^n$ being some positive
power of the related gauge coupling and
$\lam$ describing the order of magnitude of the mass of the
new heavy states. On the
contrary, the operators involving also the Higgs and/or the
quark fields have their couplings determined by the  
new Yukawa couplings and   $1/\lam^2$ of course.
Usually there is no principle to constrain the magnitude of
these couplings, which can therefore be large.
It appears therefore that the Higgs or quark involving operators,
often have a better chance to be generated by NP at an observable
level, than the purely gauge boson ones. \par

As a concrete and rather special example of these models we 
consider below the case
where the only relevant heavy degrees of freedom which need to be
considered at the $0.5-1~TeV$ region is a heavy isosinglet
colour-triplet boson $\Psi$ with hypercharge $-2/3$, and 
a Majorana
fermion $F$ with vanishing $SU(3)\times SU(2) \times U(1)$
quantum numbers. The most general CP conserving and 
renormalizable lagrangian describing this
model is obtained by adding to the SM lagrangian the interaction
\bqa
\label{LInew}
\L_{\mbox{I}} & =  & \frac{1}{2}
\left (i\ol{F} \Dsl F-M_F \ol{F} F \right )~+~D_\mu\Psi^\dagger
D^\mu\Psi - M_\Psi^2 \Psi^\dagger \Psi +2g_{\Psi}(\Psi^\dagger
\Psi) (\Phi^\dagger \Phi) \ \  \nonumber \\
& +& f (\ol{t}_R\Psi^\dagger F +\mbox{ h.c.} ) \ \ ,
\eqa
where $g_{\Psi}$ and $f$ are unknown real Yukawa couplings 
determined by NP. The masses of $\Psi$ and $F$ are generated
before the electroweak breaking and they can be naturally assumed
to be large. Below we assume for simplicity 
\bq
    M_\Psi ~\sim ~ M_f  ~\sim ~ \lam \ \ .
\eq\par

Such a model could be generated \eg\@
in a supersymmetric theory, where only  a
right stop and a singlet $B$-gaugino, identified with
$\Psi$ and $F$ respectively,  are sufficiently
light to be relevant. Thus, in such a SUSY model we assume
that all other new (s)particles, including
all Higgses except the lightest one usualy called $h$, are (say) 
above $3~TeV$ and that they are are ignored.  We also remark that in
such a context $g_\Psi$ still remains a free Yukawa parameter,
while the other Yukawa coupling $f$, is actually fixed by SUSY
to 
\bq
f ~=~-~\frac{2\sqrt{2}}{3}~g\prime \ \ .
\eq\par

After integrating out the new heavy particles, we find that
at a scale just below $\lam$, NP is described by  the
effective lagrangian \\

\bqa
\L_{NP} & = & \frac{1}{(4\pi \lam)^2} \Bigg \{
- \frac{g_s^3}{60} \O_G  
~-~  \frac{g\prime^2}{90}\ol{\O}_{DB}
~-~  \frac{g_s^2}{240}\ol{\O}_{DG} \nonumber \\
&-& g_\Psi \frac{2g\prime^2}{9} \O_{BB}
~-~ g_\Psi  \frac{ g_s^2}{12} \O_{GG} 
~+ ~  \frac{g_\Psi^3}{2} \O_{\Phi3}
~+~  \frac{g_\Psi^2}{4} \O_{\Phi2}
~+~   \frac{f^4}{12}\O_{tt}\nonumber \\
&+& f^2\, \frac{g\prime \mt}{36\sqrt{2} v} \O_{tB\Phi}
~+~ f^2\, \frac{g_s \mt}{48 \sqrt{2} v} \O_{tG\Phi}
~+~ f^2\, \frac{\mt}{6\sqrt{2} v}\, \left [4g_\Psi + 
\left (\frac{\mt}{v} \right )^2 \right ] \O_{t1}
\nonumber \\
&+& f^2 \, \frac{\mtd}{12 v^2} \O_{\Phi q}^{(1)}
~-~ f^2 \, \frac{g\prime}{36}\O_{tB}
~-~  f^2 \, \frac{g_s}{24}\O_{tG} \Bigg \}
\ \ . \ \label{LNP}
\eqa \par

We note that only 14 operators, out of the total number of 48,
are generated in this particular model. We also note that
only the Higgs or quark involving operators have their ~strengths
affected by the Yukawa couplings $g_\Psi$ and $f$.

\section{Final discussion}

In this paper we have first established the full list of 
CP conserving 
$dim=6$ and $SU(3)_c \times SU(2)\times U(1)$ gauge invariant
operators, which should describe any kind of a high scale 
New Physics induced by the Higgs and the particles most
strongly coupled to it. Taking into account the groupings implied
in the Standard Model, we identified these later particles as
being the gauge bosons and the quarks of the third
family. 
This list of NP operators consists of 14
purely bosonic ones involving photon, $Z$, $W$,
Higgs and gluon fields, as well as 34 operators containing 
in addition quarks of the third family.  \par

Subsequently, we have restricted the above list by excluding the 
"non-blind" operators, which are defined as the operators
contributing at tree level to observables measurable at LEP1/SLC
and low energy experiments.  Such "non-blind" operators are
excluded because present experiments imply values for their 
couplings  that would be too weak to produce
measurable effects at future colliders. On the other hand,
operators contributing to $Z$ peak processes
at 1-loop order, are retained, because the resulting constraints
are so mild that they leave 
room for various effects observable at higher
energies. The study of the purely gluonic 
operators $\ol{\O}_{DG}$ and $\O_G$ is postponed for the
future, since it needs somewhat detail QCD considerations. The
corresponding
constraints should come from the effect of these operators
on multijet production and also on the 
running of $\alpha_s(q^2)$. \par
 
For each of the retained operators, 
we have computed the two-body scattering
amplitudes and established the unitarity constraints. This allows
us to associate without any ambiguity a NP scale to the coupling 
constant of each of these operators. This is the scale where 
the strongest tree level amplitude approaches the unitarity bound. 
Close to this scale, we expect various manifestations of NP to 
appear, like \eg\@ creation of new particles, resonances,....\par

This result is useful in various respects. If the NP scale
is not known a priori, one can derive upper 
bounds for it, using the observability limits for
the NP couplings obtained in present experiments, 
or expected at future colliders.
For example, using the observability limits established in \cite{Bil},
one obtains that the study of anomalous 3-gauge boson couplings in
$e^+e^-\to W^+W^-$ allows to reach NP scales up to about $1.5~TeV$ at
LEP2 and $10~TeV$ at an $e^-e^+$ Linear Collider (LC) 
at $0.5~TeV$. 
With $e^+e^-\to HZ, H\gamma$, if the 
value of $m_H$  allows the Higgs to be produced, one reaches in the
study of anomalous Higgs-gauge boson couplings, scales of about $7~TeV$
at LEP2 and $20~TeV$ at LC \cite{HZGRV}. With $\gamma\gamma$ 
processes, realizable
through the laser backscattering method at LC, one should have
independent informations on these various couplings and reach NP scales
up to $20~TeV$ in $\gamma\gamma\to VV$ and $65~TeV$ in $\gamma\gamma\to
H$ \cite{NP-GLR}. Finally anomalous top quark couplings
contributing at tree level in
$e^+e^-\to t\bar t$ should be visible for NP scales less than about
$35~TeV$ at LC \cite{eettGKR}.\par

This procedure allows to make useful comparisons between different
types of operators and between different sectors in the processes 
accessible to experiment. It is free from
ambiguities in the normalizations of the coupling constants 
associated to
different types of fields (scalars, spinors, vectors,...). We can
more safely discuss the hierarchy that appears in the various
sectors, either from
theoretical reasons, or from purely experimental considerations.\par

Conversely, if one knows the NP scale from some
theoretical prejudice, one can use the unitarity constraints 
to predict the values of the NP
coupling associated to a given operator
contributing in the NP effective Lagrangian.
For example, specific New Physics schemes can fix the 
NP scale in some dynamical way, either from the mass of the new
heavy particles, or from the strength of the underlying 
interaction. \par

Finally we have also investigated which of the above operators
would be generated by NP in the $0.5-1~TeV$ region, under the
assumption that in this energy range the relevant group is
just $SU(3)\times SU(2)\times U(1)$, and that the relevant
nearby heavy particles whose integration out creates the
effective NP operators, are just scalars and fermions.
For their quantum numbers we have considered any combination
of singlets and ~doublets under isospin, and singlets and
triplets under QCD colour. This way we have identified 8
operators which are never generated. Whenever possible, we have
also mentioned the conditions determining the appearance or not
of some of the operators that can in principle be 
generated. A specific model was presented in which
NP is described by only 14 of the above operators.\par     

We conclude that experimental limits on NP effects from present and 
future data when expressed in terms of NP scales and compared with 
theoretical landscapes such as those suggested by these dynamical models
should be instructive when looking for hints about the origin and the
basic sturcture of NP.

 \par

%\newpage


\begin{thebibliography}{99}



\bibitem{SM1} M. Veltman, \np{B123}{1977}{89}; M.E. Peskin and T.
Takeuchi, \prl{65}{1990}{964}; G. Altarelli and R. Barbieri,
\pl{B253}{1991}{161}.
% 
\bibitem{SM2}A. Blondel, ICHEP96 (plenary talk), Warsaw July
1996. W. Hollik, KA-TP-19-1996, hep-ph/9608325.
%
\bibitem{NPgen}G.J. Gounaris and F.M. Renard, 
\zp{C59}{1993}{133}; 
J.D. Wells, C. Kolda and G.L. Kane
\pl{B338}{1994}{219}; 
 C.T. Hill and X. Zhang, \pr{D51}{1995}{3563};
T.G. Rizzo, \pr{D51}{1995}{3811}; 
H. Georgi, L. Kaplan, D. Morin and A. Schenk,
\pr{D51}{1995}{3888}; Z. Zhang and B.-L. Young,
\pr{D51}{1995}{6584}; E. Ma and D. Eng, \pr{D53}{1996}{255};
%
\bibitem{Chano} M.S.Chanowitz and M.K.Gaillard, \np{B261}{1985}{379};
    M.S.Chanowitz, Ann.Rev.Nucl.Part.Sci.38(1988)323.
%
\bibitem{NP-GLR}
G.J. Gounaris, J. Layssac and F.M. Renard, \zp{C69}{1996}{505}.
G.J. Gounaris and F.M. Renard, \zp{C69}{1996}{513}.

\bibitem{Appel} T.Appelquist and G-H.Wu \pr{D48}{1993}{3235}.
%
\bibitem{NPnonlin}H.-J. He, Y.-P. Kuang and C.-P. Yuan, 
DESY 95-252, hep-ph/9604309.
%
\bibitem{Buchmuller} W. Buchm\"{u}ller and D. Wyler,
\np{B268}{1986}{621}; C.J.C. Burgess and H.J. Schnitzer,
\np{B228}{1983}{454}; C.N. Leung, S.T. Love and S. Rao
\zp{C31}{1986}{433}. C. Arzt, M.B. Einhorn and J. Wudka,
\np{B433}{1995}{41}. 
%
\bibitem{Hag-boson}
K.Hagiwara et al, \pl{B283}{1992}{353};\pr{D48}{1993}{2182}.
% 
\bibitem{topGRV1} G. Gounaris, F.M.Renard and C.Verzegnassi, 
\pr{D52}{1995}{451}.
%
\bibitem{top-other}D. Atwood, A. Kagan and T.G. Rizzo 
\pr{D52}{6264}{1994}.
X. Zhang and B.-L. Young \pr{D51}{1995}{6564}. 
T. Han, R.D. Peccei and X. Zhang, \np{B454}{1995}{527}.
T. G. Rizzo, \pr{D53}{1996}{6218}.
P. Haberl, O. Nachtman and A. Wilch, \pr{D53}{1996}{4875}. 
%
\bibitem{BLYoungH} M. Hosch,  K. Whisnant and B.-L. Young, 
AMES-HET 96-04.
%
\bibitem{Georgi}H. Georgi, \np{B361}{1991}{339}.
%
\bibitem{unit1} G.J. Gounaris, J. Layssac and F.M. Renard,
\pl{B332}{1994}{146}; G.J. Gounaris, J. Layssac, J.E. Paschalis
and F.M. Renard,  \zp{C66}{1995}{619}.
% 
\bibitem{Tsirigoti2} G.J. Gounaris, F.M. Renard and G. Tsirigoti,
\pl{B350}{1995}{212}. 
%
\bibitem{HZGRV} G.Gounaris, F.M.Renard and N.D.Vlachos
Nucl.Phys. B459(1996)51.
%
\bibitem{Altarelli} G. Altarelli, R. Barbieri and F.
Caravaglios \pl{B314}{1993}{357}.
%
\bibitem{DeR} A. De R\'{u}jula \etal~, \np{B384}{1992}{3}.
%
\bibitem{BLYoungOt1}K. Whisnant, B.-L. Young and X. Zhang, 
\pr{D52}{1995}{3115}. Bl-L. Young, Talk at Int. Symp. on Heavy
Flavour and Electroeak Theory, August 1995, Beijing, China,
hep-ph/9511282.
%
\bibitem{Tsirigoti1}G.J. Gounaris, F.M. Renard and G. Tsirigoti,
 \pl{B338}{1994}{51}.
%
\bibitem{Arzt}C. Arzt, M.B. Einhorn and J. Wudka,
\np{B433}{1995}{41}.
%
\bibitem{Papadamou-dyn} G.J. Gounaris, D.T. Papadamou and F.M.
Renard, PM/96-31, THES-TP 96/10, hep-ph/9611224. 
%
\bibitem{Bil} M. Bilenky, J.L. Kneur, F.M. Renard and
D. Schildknecht, \np{B409}{1993}{22} and
{\bf{B419}} (1994) 240.
% 
\bibitem{eettGKR} G.J.Gounaris, M. Kuroda and F.M.Renard,
\pr{D54}{1996}{6861}.  







 
\end{thebibliography}
\end{document}